# Interplay of Water and a Supramolecular Capsule for Catalysis of Reductive Elimination Reaction from Gold

Valerie Vaissier Welborn[1,2,3*], Wan-Lu Li[1,2,3], Teresa Head-Gordon[1-5†]

[1]Kenneth S. Pitzer Center for Theoretical Chemistry
[2]Department of Chemistry, University of California, Berkeley
[3]Chemical Sciences Division, Lawrence Berkeley National Labs, Berkeley
[4]Department of Chemical and Biomolecular Engineering and
[5]Department of Bioengineering, University of California, Berkeley
E-mail: thg@berkeley.edu

**ABSTRACT**

Supramolecular assemblies have gained tremendous attention due to their apparent ability to catalyze reactions with the efficiencies of natural enzymes. Using Born-Oppenheimer molecular dynamics and density functional theory, we identify the origin of the catalytic power of the supramolecular assembly $Ga_4L_6^{12-}$ on the reductive elimination reaction from gold complexes and their similarity to enzymes. By comparing the catalyzed and uncatalyzed reaction in explicit solvent to identify the reaction free energies of the reactants, transition states, and products, we determine that a catalytic moiety – an encapsulated water molecule – generates electric fields that contribute significant reduction in the activation free energy. Although this is unlike the biomimetic scenario of catalysis through direct host-guest interactions, the nanocage host preconditions the transition state for greater sensitivity to electric field projections onto the breaking carbon bonds to complete the reductive elimination reaction with greater catalytic efficiency. However it is also shown that the nanocage poorly organizes the interfacial water, which in turn creates electric fields that misalign with the breaking bonds of the substrate, thus identifying new opportunities for catalytic design improvements in nanocage assemblies.

[*]*Department of Chemistry, Virginia Tech University, Blacksburg, VA 24061*
[†]Correspondence to: thg@berkeley.edu

**INTRODUCTION**

Inspired by the ultimate enzyme catalyst[1, 2], supramolecular catalytic systems seek to be biomimetic[3, 4, 5] for features such as presence of an active site[6] with optimized non-covalent interactions with the substrate,[7] electrostatic preorganization that eliminates the reorganization cost paid by the uncatalyzed reaction,[8, 9] as well as desolvation and dynamical effects that are relevant for the complete catalytic cycle.[2, 10, 11] In turn, the weak intermolecular interactions that govern supramolecular assemblies offer the undeniable advantage of easy functionalization, reversibility, and fast self-assembly[5, 12, 13] that overcome limitations of a delicate and more difficult redesign of an enzyme scaffold[14]. Supramolecular catalysts have the potential to revolutionize the chemical industry by allowing simpler and more flexible reaction pathways that offers lower cost, reduces the creation of undesired byproducts[15, 16, 17, 18], operates in a broad range of conditions.[19, 20], and is compatible with renewable and sustainable man-made chemistry.[18, 21]

Nanocapsule or cage-like supramolecular catalysts have attracted a lot of attention due to their perceived similarities to enzymes and their remarkable efficiencies.[3, 10, 11, 18, 22] Theoretical calculations have become indispensable to thoroughly analyze the role of the encapsulation during the catalytic process for reactions including Diels-Alder[23, 24], ester hydrolysis[25], decarboxylation inside $\beta$-cyclodextrin[26] and in the so-called softball complex[27]. The catalytic mechanism of nanocage supramolecular systems have been categorized as (i) encapsulation of a catalytic moiety, thereby shielding the reaction from undesired bulk side reactions[28] and (ii) encapsulating only reactants and to rely on host-guest interactions to promote the reaction as do enzymes[28, 29, 30]. What are thought to be the most notable examples of the latter are the metal-ligand assemblies $M_6L_4$ introduced by Fujita et al[4, 29, 31] which can tune the Diels Alder reaction towards the formation of new products, and the $M_4L_6$ assembly introduced by Raymond and co-workers (Figure 1, top), that catalyzes a number of reactions including Nazarov cyclization of dienol substrates as well as aza-cope rearrangements of cationic enammoniums, with enzymatic efficiencies.[4, 22, 32, 33, 34]

Relevant to this work, $Ga_4L_6^{12-}$ has been proven to accelerate the alkyl-alkyl reductive elimination from gold(III) complexes by five order of magnitude in a methanol/water solvent mixture.[35, 36] Subsequent experimental studies have revealed that the nanocage catalyzed reaction obeys Michaelis-Menten kinetics, and demonstrating that the $Ga_4L_6^{12-}$ capsule creates a microenvironment that preferentially binds a cationic intermediate as the substrate (Figure 1, bottom).[35, 37] Our group has shown that the total activation *potential* energy of the reaction from this cationic intermediate is lowered by the electrostatic environment emanating from the $Ga_4L_6^{12-}$ system relative to that of a $Si_4L_6^{8-}$ capsule, which is consistent with catalytic trends observed experimentally.[1] A recent study by Ujaque and co-workers have proposed that the origin of the catalytic behavior in pure methanol arises from two factors: (i) encapsulating the gold complex inside the $Ga_4L_6^{12-}$ nanocage accounting for the interaction and thermal terms of the overall process, and (ii) removing explicit methanol microsolvation around the encapsulated gold complex to get better fits within the cavity.[38] But what has not yet been characterized is the aqueous solvent component and its role in driving the catalytic effect.

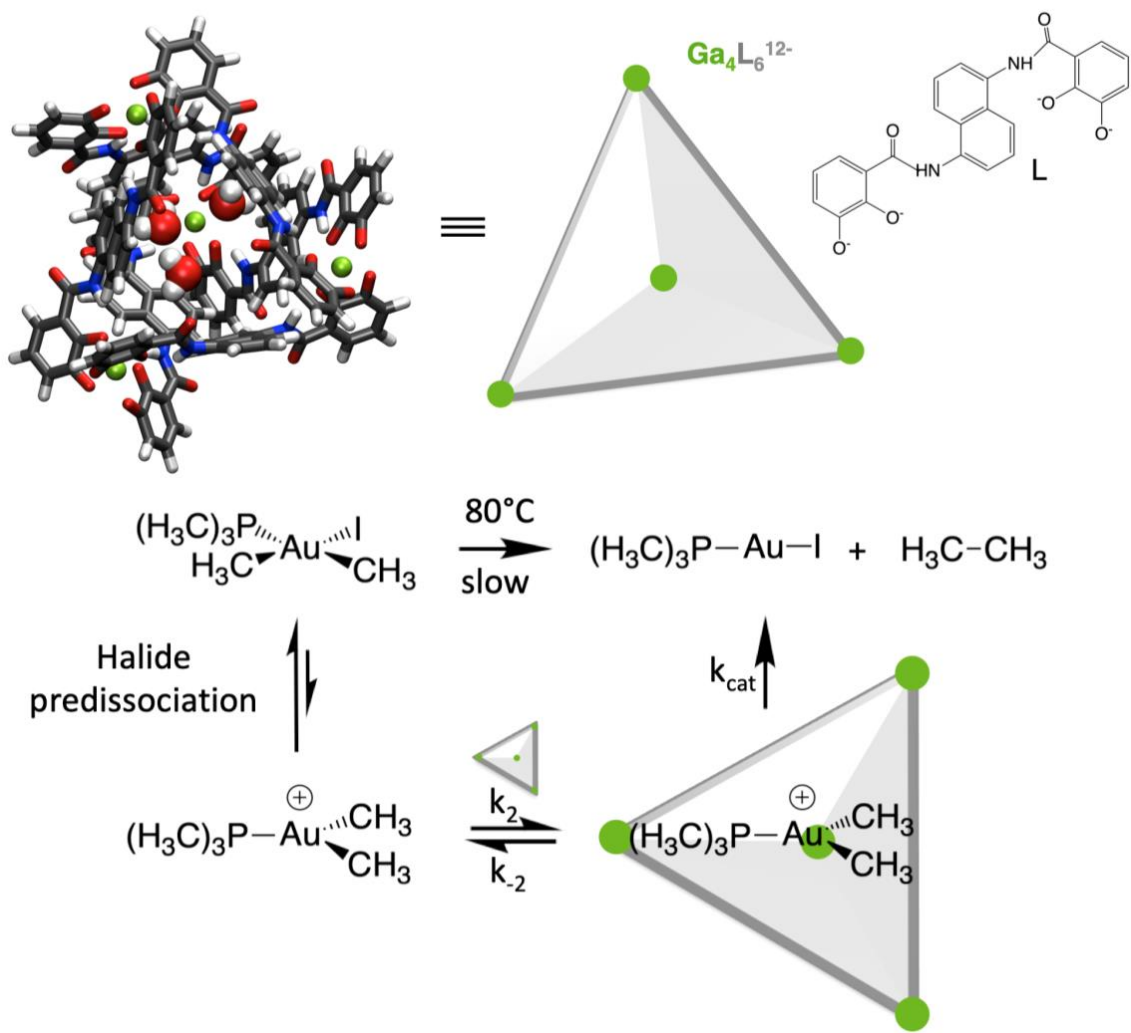

Figure 1. Supramolecular catalyst $Ga_4L_6^{12-}$ for the reductive elimination from gold(III) complexes. Top: the $Ga_4L_6^{12-}$ (L=N,N'-bis(2,3-dihydroxybenzoyl)-1,5-diaminonaphtalene) tetrahedral assembly. Bottom: the substrate trialkylphosphine(dimethyl)gold iodide ($P(CH_3)_3(CH_3)_2AuI$) is in equilibrium with $P(CH_3)_3(CH_3)_2Au_+$, and the reaction occurring in the nanocage binds the positively charged unhalogenated form as proposed in Ref[35]. Color key: carbon=gray, nitrogen=blue, hydrogen=white, oxygen=red, gallium=green.

Analysis of the reactivity of nanocage or nanoconfined complexes by theoretical approaches can unravel many novel physicochemical properties of the catalytic systems under consideration.39 More recent research has now moved toward more systematically using explicit solvent under periodic boundary via ab initio molecular dynamics (AIMD) for catalytic reactions, especially for organometallic systems.40, 41 Furthermore, the free energy is an important missing factor in the previous computational studies of alkyl-alkyl reductive elimination from gold(III) complexes in solution and in the metalloenzyme.1 For example, one of the major factors in the cycloaddition reaction acceleration inside a cucurbit[6]uril host is the reduction of the entropic component of bringing reactants together at the reaction barrier.42, 43 Himo and co-workers predicted that the dominant contributor to the rate acceleration is the entropic effect along with destabilization of the reactant in the presence of resorcinarene-based host.44 Although water has been shown experimentally to influence the $Ga_4L_6^{12-}$ nanocage's ability to catalyze acid-, base- and water-mediated proton transfer33, the role of water has not been elucidated for reactions that do not involve proton transfer.

Here we seek to quantify the reaction mechanism of the $Ga_4L_6^{12-}$ catalyzed alkyl-alkyl reductive elimination using ab initio molecular dynamics (AIMD) of the nanocage in explicit water solvent, and further analyzed with metadynamics and committor analysis to determine the free energy surface. We find that the interfacial water solvent surrounding the $Ga_4L_6^{12-}$ nanocapsule generates electric fields that oppose the catalytic acceleration, unlike enzymes whose scaffold is evolutionary designed to minimize the reorganization energy.2, 14 Furthermore we find that the $Ga_4L_6^{12-}$ nanocapsule is not in fact biomimetic of an enzyme active site governed by host-guest interactions, but rather that the nanocage creates a catalytic moiety- a primary water molecule- that generates bond dipole-field interactions that preferentially stabilize the transition state, thereby overcoming the poor interfacial solvent organization to accelerate the alkyl-alkyl reductive elimination reaction. Together, these results have identified new biomimetic design strategies to increase the catalytic power of supramolecular assemblies in the future.

**RESULTS**

**Free energy of alkyl-alkyl reductive elimination with and without $Ga_4L_6^{12-}$**

To rationalize the role of the nanocage construct on the evolution of the reductive elimination, we compare the reaction path of both the catalyzed and uncatalyzed reactions using ab initio metadynamics and a frozen string method to determine reactants, products, and transition states (see Methods). To accelerate the exploration of the free energy landscape, we picked two coordinates as collective variables and calculate the free energy pathway in the reduced two-dimensional space. In this study, our choice of (i) the distance between the carbon atoms of the leaving methyl groups (methyl-methyl distance) and (ii) the coordination number between the leaving carbons and the gold was found to yield a correct transition state ensemble as subsequently confirmed with a committor analysis using unconstrained AIMD trajectories in the complete higher dimensional space of the reaction (see Methods).

The free energy barrier calculated from these minimum energy paths is 37 and 33 kcal mol$^{-1}$ (with (Figure S1) and without iodide (Figure 2), respectively) for the uncatalyzed reaction and 24 kcal mol$^{-1}$ for the catalyzed reaction. We note that for the pre-equilibrium step (Figure 1), the dominant species in aqueous solution is the halogenated form, but the rate law for catalysis is dependent on the unhalogenated substrate.[36] Using a simple transition state theory (TST) relationship, $exp(\Delta G^{\ddagger}/k_b T)$, this would correspond to a rate acceleration of 3.3 x 10$^7$, in reasonable agreement with 5.0 x 10$^5$ to 2.5 x 10$^6$ (depending on halide ligand) that was determined experimentally.[36] The quantitative values for the activation free energy may be limited by the TST assumption, or perhaps the level of DFT theory, although the more advanced range-separated hybrid version of the B97M functional complemented with a triple zeta basis set has been well-validated on bulk water[45, 46], is likely to describe the catalytic system accurately. Alternatively, it may also stem from the fact that the original experiment was carried out in a methanol/water mixture whereas we are considering pure water solvent. The presence of less polarizable methanol molecules would diminish the electrostatic interactions around the gold complex.

**Origin of the catalytic power of Ga$_4$L$_6$$^{12-}$.**

To better understand how the Ga$_4$L$_6$$^{12-}$ nanocage provides a total of ~9 kcal mol$^{-1}$ reduction in the activation energy barrier, we selected snapshots characteristic of the reactant and transition states from the AIMD trajectory (Figure 2). Our group has previously analyzed the electrostatic environment of enzyme active sites to show they create large electric fields that are well-aligned with reactive bonds, and act as an important contributor to transition state stabilization as well as reactant state destabilization.[1, 2] Therefore we calculated the electrostatic part of the energy barrier by computing electric fields from different system components, and their contributions to the activation energy barrier. This helps to quantify the catalytic role of the nanocage and encapsulated water molecule vs. that of the greater bulk water environment of the uncatalyzed reaction.

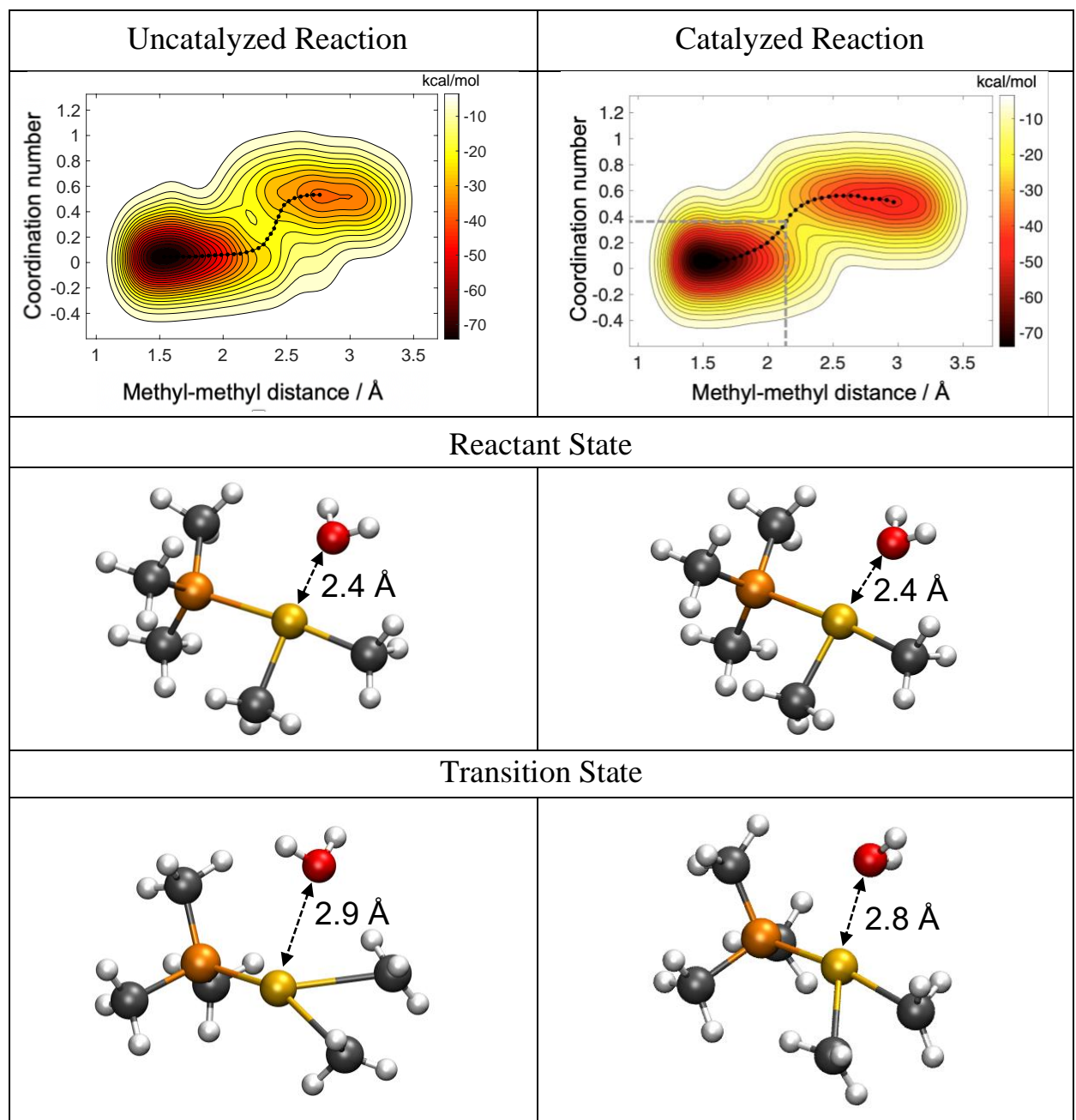

Figure 2. Free energy landscape obtained from ab initio metadynamics. The uncatalyzed reaction on the left describes the evolution of the unhalogenated compound P(CH$_3$)$_3$(CH$_3$)$_2$Au$_+$ in water. The catalyzed reaction on the right describes the evolution of the encapsulated cation P(CH$_3$)$_3$(CH$_3$)$_2$Au$_+$ in water. As an additional reference, we provide in Supporting Information the landscape for the cation P(CH$_3$)$_3$(CH$_3$)$_2$AuI in water in the absence of the cage (Figure S1). Color key: carbon=gray, phosphorous=orange, gold=yellow, hydrogen=white, oxygen=red.

To further quantify the effect of the electric fields, we estimated the electrostatic free energy of stabilization of the transition state, $\Delta G^{elec}$, defined as follows:

$$\Delta G^{elec} = \sum_i -0.048(\boldsymbol{\mu}^i_{TS}\cdot \mathbf{E}^i_{TS} - \boldsymbol{\mu}^i_{RS}\cdot \mathbf{E}^i_{RS}) \quad (1)$$

where the summation is over the number of reactive bonds considered, $\vec{\mu}^i_X$ is the bond dipole moment and $\vec{E}^i_X$ the electric field in state X. Since both the magnitude and the orientation of electric fields are of importance, we projected the fields onto the two bonds that change the most during the reaction, identified as the two gold-methyl bonds as shown in Table 1.

Table 1: Electric fields ($E_1$ and $E_2$ respectively) and corresponding free energies ($\Delta G^\dagger$) projected onto the two bonds that break during the reaction. The electric fields are given by the contribution from bulk water, a vicinal water, and from the nanocage for the reactant (RS) and transition state (TS) of the uncatalyzed and catalyzed reaction. Positive fields are defined in the opposite direction of the flow of electrons and contribute to stabilizing electrostatic effects. The bond dipoles ($\vec{\mu}^1_{RS}, \vec{\mu}^2_{RS}, \vec{\mu}^1_{TS}, \vec{\mu}^2_{TS}$) were computed from the partial charges on the gold and carbon atoms, and using the bond length $d_{Au-Ci}$ as shown in Figure S2 and Table S1. The unit conversion factor for free energy from the projected electric field on the bond dipole in kcal/mol is 0.048. Color key: carbon=gray, phosphorous=orange, gold=yellow, hydrogen=white, oxygen=red.

| Electric fields | | $E_1$ / MV cm$^{-1}$ | | | $E_2$ / MV cm$^{-1}$ | | |
|---|---|---|---|---|---|---|---|
| | | Bulk Water | Complexed water | Cage | Bulk water | Complexed water | Cage |
| RS | Uncatalyzed reaction | –21.49 | 59.73 | N/A | –12.75 | –14.89 | N/A |
| | Catalyzed reaction | –6.72 | 40.81 | –0.21 | –6.87 | –7.97 | –14.13 |
| TS | Uncatalyzed reaction | –4.55 | 9.63 | N/A | –15.81 | –21.2 | N/A |
| | Catalyzed reaction | –51.63 | 19.30 | 27.27 | –31.37 | 22.27 | 9.95 |

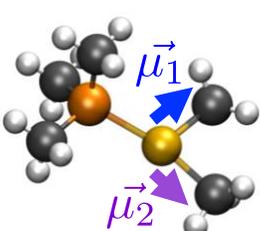

| | $\boldsymbol{\mu}^1_{RS}$ | $\boldsymbol{\mu}^2_{RS}$ | $\boldsymbol{\mu}^1_{TS}$ | $\boldsymbol{\mu}^2_{TS}$ | Bulk water | Complexed water | Cage |
|---|---|---|---|---|---|---|---|
| Uncatalyzed reaction | –6.15 | –2.83 | –1.26 | 3.67 | 10.58 | –11.29 | N/A |
| Catalyzed reaction | –6.15 | –2.83 | 2.05 | 0.71 | 9.02 | –13.62 | –1.04 |

$$\Delta G^{elec} = -\sum_i (\boldsymbol{\mu}^i_{TS}\cdot \mathbf{E}^i_{TS} - \boldsymbol{\mu}^i_{RS}\cdot \mathbf{E}^i_{RS})$$

**DISCUSSION**

When comparing the catalyzed to uncatalyzed reaction, we see that the electrostatics alone provide ~5 kcal mol$^{-1}$ of the 9 kcal mol$^{-1}$ transition state stabilization. However the nanocage itself, although producing large electric fields consistent with its high negative charge, plays a mixed role in the catalytic effect on the carbon reductive elimination reaction from gold. We first note that, unlike enzymes whose scaffold is evolutionary designed to minimize the reorganization energy[2, 14], the nanocage poorly organizes the interfacial/bulk water which in turn creates electric fields that misalign with the breaking bonds of the substrate (Table 1). But relative to the uncatalyzed reaction, the nanocage does contribute ~50% reduction in the activation free energy, both directly through host-guest interactions, and indirectly through partial reorganization of the interfacial water near (but outside) the nanocage to be less detrimental to the reaction.

However, the remaining ~50% of the transition free energy stabilization comes from a single complexed water encapsulated with the reactants in the cage. In this context, the role of the nanocage is to generate a micro-environment in which this phenomenon is possible, which contrasts from previous speculations that put forward host-guest interactions as the main catalytic process.[28, 29, 30] The nanocage does play another implicit role for catalysis since the transition state structure is different in the nanocage when compared to the bulk, and in turn contributes to changes in the bond dipoles. In other words, the nanocage increases the system's sensitivity to the electric fields, although the true catalytic effect comes from the isolated water molecule(s) within the cage.

In conclusion, the theory presented here provides new insights into the catalytic power of the cage-like supramolecular catalyst $Ga_4L_6^{12-}$. For the alkyl-alkyl reductive elimination from gold(III) complexes, we show here that the two traditional categories to explain their catalytic process - i.e. cage-like compounds that encapsulate a catalytic moiety and the ones that use host-guest mechanisms - are actually not so easily separable. The $Ga_4L_6^{12-}$ nanocage both stabilizes the catalytic reactant through loss of a halide ligand, and preconditions the transition state for greater sensitivity to electric field projections onto the breaking carbon bonds, but also hosts additional water molecules, of which one complexed guest water serves as a strong catalytic player. At the same time the interfacial water is found to be highly detrimental to transition state stabilization, thereby identifying catalytic design opportunities for supramolecular assemblies such as $Ga_4L_6^{12-}$ to further accelerate the reductive elimination reaction from gold complexes.

DATA AVAILABILITY

The datasets generated during and/or analysed during the current study are available from the corresponding author on reasonable request

METHODS

DFT calculations

All calculations presented in this paper (geometry optimization, molecular dynamics, metadynamics, energy calculations) were performed with Density Functional Theory (DFT) using the dispersion corrected meta-generalized gradient approximation (GGA) functional B97M-V[47, 48] in combination with a DZVP basis set optimized for multigrid integration[49] as implemented in the CP2K software package.[50, 51] In all cases, we used periodic boundary conditions, 5 grids and a cutoff of 400 Ry.

Starting geometries

The starting geometry for the catalyzed reaction is the cation gold complex encapsulated in the cage. This was built by positioning the vacuum optimized cation geometry in the capsule minimizing the root-mean-square-displacement (RMSD) with the X-ray structure of bis(trimethylphosphine) gold cation in $Ga_4L_6^{12-}$. The overall structure was further optimized with DFT. The starting geometry for the uncatalyzed reaction is the vacuum optimized gold complex. Both of these structures were then solvated using Gromacs with a pre-equilibrated water box of size 30Å x 30Å x 30Å for the encapsulated gold complex and 19Å x 19Å x 19Å for the reference reaction (without the cage). To maintain charge neutrality, potassium counter ions were also included at the positions provided in the X-ray resolved structure[35] for the encapsulated system. We ran an additional 5ps ab initio molecular dynamics simulation (298K, 0.5 fs timestep) to further equilibrate the structures.

Ab initio metadynamics.

Using these equilibrated solvated structures, we then ran well-tempered single walker metadynamics[52, 53] as implemented in the CP2K package. To reduce the dimensions of the space to explore, we picked two collective variables that best describe the evolution of the reaction: (i) the distance between the carbon of the two leaving methyl groups and (ii) the coordination number ($CN_{C\text{-}Au}$) between the gold atom and the two carbons of the leaving methyl groups defined as follows:

$$CN_{C\text{-}Au} = \frac{1}{2} \sum_{i \in \{1,2\}} \frac{1 - (\frac{r_{C_i\text{-}Au}}{R_0})^8}{1 - (\frac{r_{C_i\text{-}Au}}{R_0})^{14}} \quad (2)$$

Where $r_{C_i\text{-}Au}$ ($R_0$) is the instantaneous (equilibrium) distance between the gold and carbon atoms. The choice of these coordinates was guided by our previous study for which we computed the geometry in vacuum of the reactant, transition state and product of the alkyl-alkyl reductive elimination reaction. It is also worth noting that, given the nature of the system, other candidates for collective variables (such as angles or dihedrals) would likely depend on either if not both the gold-methyl coordination number or the methyl-methyl distance.

In this metadynamics scheme, Gaussian functions of height 0.005 Ha were deposited at least every 30 steps (with a time step of 0.5 fs) along the trajectory in the reduced space. This introduces a history dependent bias that pushes the system towards areas of the landscape that would otherwise be hard to reach (such as the crossing between reactant and product wells). For both the catalyzed (with cage) and uncatalyzed (without cage) reactions, this process was run until the barrier was crossed at least 3 times, gathering over 50ps of metadynamics. Free energy surfaces were then created using the sub-program graph within the CP2K package. This tool reads in the information about the added Gaussian functions such as position, height and width and compute the corresponding unbiased energy landscape. From these, minimum energy paths were calculated using the zero temperature string method of Maragliano et al.[54] The procedure was performed in Matlab using a 30 point string and 3000 optimization steps.

To rationalize the role of the cage on the reduction of the reaction energy barrier, a set of geometries representative of the reactant and transition states were extracted. For the reactant state, we selected one structure every 10 fs of the molecular dynamics trajectory for 2 ps (200 structures total). For the transition state, we searched for a few snapshots within the metadynamics trajectory that corresponded to the values of the collective variables identified as transition state by the string method. We then tested and refined our choice by performing a committer analysis until our selected geometries would fall in both the reactant and product equilibrium well. For the catalyzed reaction, we found 3 snapshots that displayed a 57% (43%) commitment to the product (reactant) state. These were situated at (2.1, 0.35) in the collective variable space, very close to the initial guess obtained by the string method (2.1, 0.36). Similarly, for the uncatalyzed reaction, we found 2 snapshots that displayed an early 38% (62%) commitment to the product (reactant) state, at (2.6, 0.24) in the collective variable space compared to the initial guess of (2.3,0.24). An ensemble was then generated by gathering all geometries that fell within +/0.05A in methyl-methyl distance and +/-0.005 in coordination number to (2.1, 0.35) for the catalyzed and (2.6, 0.24) for the uncatalyzed reaction. This represents about 45 geometries for each transition state ensemble for complete committor analysis statistics, thereby confirming that the transition state was reliably found.

**Electric fields.**

The derivative of the electrostatic potential were obtained as direct output of CP2K, and the electric field was then projected onto the two bonds of the substrate that are most changed during the evolution of the reaction, namely the two gold-carbon of the leaving methyl group bonds (see Figure 5). The free energy state functions were obtained from this electric field projection through a model of the bond dipoles that were computed using the Density Derived Atomic Point Charge (DDAPC)[55] scheme that accounts for the multigrid integration of CP2K. The charges, bond lengths and details of these calculations are given in Supporting Information.

**ACKNOWLEDGEMENTS**

This work was supported by the Director, Office of Science, Office of Basic Energy Sciences CPIMS program, Chemical Sciences Division of the U.S. Department of Energy under Contract No. DE-10AC02-05CH11231. This research used computational resources of the National Energy Research Scientific Computing Center, a DOE Office of Science User Facility supported by the Office of Science of the U.S. Department of Energy under Contract No. DE-AC02-05CH11231, under an ASCR Leadership Computing Challenge (ALCC) award.

**Conflict of interest.** The authors declare that they have no conflict of interest.

**Author contribution.** THG and V.W. conceived the scientific content and direction, V.W., W.L., and THG wrote the manuscript, and V.W. and W.L. created the Figures. All authors contributed data and insights, discussed and edited the manuscript.